\documentclass[article,twocolumn,pra,showpacs]{revtex4-1}

\usepackage{amsmath,xcolor}
\usepackage{amsfonts}
\usepackage{amssymb}
\usepackage{graphicx}
\usepackage{srcltx}
\usepackage{mathcomp} 
\usepackage{ulem} 

\newcommand{\beq}{\begin{equation}}
\newcommand{\eeq}{\end{equation}}
\newcommand{\ket} [1] {|#1\rangle}
\newcommand{\bra} [1] {\langle#1|}

\newcommand{\dens}[1]{|#1 \rangle\langle #1|}

\begin{document}

\title{Device-independent entanglement-based Bennett 1992 protocol}

%
\author{Marco Lucamarini$^1$}
\author{Giuseppe Vallone$^{2,3}$}
\author{Ilaria Gianani$^2$}
\author{{ Paolo Mataloni$^{2,5}$}}
\author{{ Giovanni Di Giuseppe$^{1,4}$}}
%

\affiliation{\vspace{0.2cm}$^1$Scuola di Scienze e Tecnologie,
Divisione di Fisica, I-62032 Camerino (MC), Italy}
\affiliation{$^2$Dipartimento di Fisica, Universit\`{a} Sapienza di Roma, I-00185 Roma, Italy}
\affiliation{$^3$Department of Information Engineering, University of Padova, I-35131 Padova, Italy}
\affiliation{{ $^4$ CriptoCam s.r.l., Via Madonna delle Carceri 9, I-62032 Camerino (MC), Italy}}
\affiliation{$^5$Istituto Nazionale di Ottica, Consiglio Nazionale
delle Ricerche (INO-CNR), L.go E. Fermi 6, I-50125 Firenze, Italy}
%

%
\begin{abstract}
\noindent In this paper we set forth a novel connection between
the Bennett 1992 protocol and a Bell inequality. This allows us to
extend the usual prepare-and-measure protocol to its
entanglement-based formulation. We exploit a recent result in the
framework of device-independent quantum key distribution to provide a
simple, model-independent, security proof for the new protocol.
The minimum efficiency required for a practical implementation of
the scheme is the lowest reported to date.
\end{abstract}
%



\maketitle

\section{Introduction}
\label{sec:1_intro}

On 1992 Charles Bennett introduced his famous minimal-state
protocol for Quantum Key Distribution (QKD), named after him
``B92''~\cite{Ben92}. It makes use of two nonorthogonal quantum
states  to convey one bit of information from a transmitting user (Alice) to a receiving
user (Bob).

The single-photon B92 protocol was proven unconditional
secure in~\cite{TKI03,TL04}. Its main problem is the unambiguous
state discrimination (USD)~\cite{Che98} attack, initially
discussed in~\cite{Ben92} and later on analyzed in~\cite{TL04},
which dramatically reduces its tolerance to the losses of the
communication channel and, by consequence, its applicability to a
practical scenario.
In this respect, a novel version of { the single-photon} B92
{ protocol,} robust against the USD { attack,} has been
recently introduced in~\cite{LDT09}. { It} exploits two
additional states in Alice's preparation, called ``uninformative
states'', to let the users detect a USD attack. {For this
reason it was called} ``us-B92''~\cite{LDT09}.

At variance with the more popular BB84 protocol~\cite{BB84}, the
B92 does not allow for an entanglement-based realization and can
only be implemented in a prepare-and-measure (PM) configuration.
This means that in order to guarantee its unconditional security
it is necessary to enclose the source of photons in Alice's
territory, well shielded against a malicious presence (Eve)
eavesdropping on the quantum channel. So, for instance, it is not
possible to place the light source in the middle of Alice and Bob
in order to increase the maximum working
distance~\cite{WZY02,Ma2007} of a B92-based QKD session.
{ The us-B92 protocol follows the same fate as the B92, and is
only PM as well. However} we realized that it admits a
straightforward  extension to a description and implementation which are based on
the entanglement, called
"ent-B92" henceforth.

{Several security proofs of B92 rely on entanglement distillation.
However the entanglement is used only as a mathematical tool to demonstrate security, 
but the physical realization was always based on PM scheme. 
Here we propose to use the entanglement as the physical resource to realize the cryptographic protocol. 
In this case it is possible to implement the protocol with the entanglement source placed in an untrusted location, between Alice and Bob.
The security proofs of the standard B92 \cite{TKI03,LDT09} and of the us-B92 \cite{LDT09} are
 valid only assuming that the entangled source is shielded in Alice side.
 In our protocol even if the entanglement source is under Eve's control, we will demonstrate that it is possible to prove its security by
connecting it with a particular form of Bell
inequality~\cite{Bell64}}, put forward for the first time by
Clauser and Horne in 1974~\cite{CH74} and later on adapted to
non-maximally entangled states by Eberhard~\cite{Ebe93}. 
This connection, besides giving a new physical insight into
a long-standing protocol like the B92, allows us to provide a
simple security proof for the { new protocol, which exploits}
a recent work by Masanes \underline{\textit{et al.}}~\cite{MPA11}
in the frame of device-independent (DI) QKD.

Hence, the ent-B92 is proven u	e regardless of
the particular implementation of the protocol. The security
proof employed is totally different from the standard
one~\cite{TKI03,TL04}, which is based on the approach described
in~\cite{Lo1999,SP00}. Notwithstanding, the obtained security
threshold is remarkably close to the one given in the literature,
thus confirming the B92 state of the art. In addition to this, we
managed to exploit the novel protocol and its security proof to
decrease considerably the minimum detection efficiency for a
possible realization of a DI-QKD, from 92.4\% reported
in~\cite{PAB+09} to 75\% of our approach. 

\section{ent-B92 protocol}
\label{sec:ent-B92}

Here we introduce the entanglement-based ent-B92  protocol that can be reduced to the B92~\cite{Ben92} or us-B92~\cite{LDT09} protocols.
Let's suppose that Alice and Bob share the following non-maximally
entangled state:
\beq
\label{eq:entangled}
\begin{aligned}
 |\Phi\rangle_{AB} =&
(|0_z\rangle_{A}|\varphi_0\rangle_{B}+|1_z\rangle_{A}|\varphi_1\rangle_{B})
/\sqrt2
\\
=&
\beta|0_x\rangle_{A}|0_x\rangle_{B}+\alpha|1_x\rangle_{A}|1_x\rangle_{B},
\end{aligned}
\eeq
where 
\begin{equation}\label{Al-qubit}
    \ket{\varphi_{j}} = \beta \ket{ 0_{x}} +( -1)^{j}\alpha
\ket{1_{x}},
\end{equation}
$\left\{ \left\vert 0_{z}\right\rangle ,\left\vert
1_{z}\right\rangle \right\} $ ($\left\{ \left\vert 0_{x}\right\rangle ,\left\vert
1_{x}\right\rangle \right\} $) are the eigenstates of the
$\textbf{Z}$ ($\textbf{X}$) basis with 
$\left\vert j_{x}\right\rangle =[\ket{0_{z}} +(-1)^{j}\ket{1_z}] /\sqrt{2}$,
$\beta =\cos\frac{\theta}{2}$, $\alpha =\sin\frac{\theta}{2}$ and
 $\theta\in(0,\pi/2)$. The state $ |\Phi\rangle_{AB}$ has been used in~\cite{Hardy93} and
\cite{Ebe93} { to propose novel tests of local realism} and
{ is} routinely implemented in laboratory~\cite{whit99prl,
cine04pra, vall07pra, vall11pra}.

If Alice measures along the $\textbf{Z}$ basis, she will project
Bob's state in either $|\varphi_0\rangle$ or $|\varphi_1\rangle$,
with equal probabilities.  {This
was at the basis of the B92 unconditional security proof given in Ref.~\cite{TKI03}. However,
it was accompanied by the further assumption that the source of
the entangled photons must be placed in Alice's secure location,
hence unreachable to Eve. This assumption is very reasonable if one is
interested to use a prepare-and-measure
(PM) protocol. When the actual protocol is entanglement-based,
such an assumption should be avoided. Here we  show that the protocol
is still  secure even when the light source is placed midway between
the users.
This ent-B92 scheme can be seen as the entanglement version of the}
 PM B92 scheme~\cite{Ben92} in which Alice prepares and sends to Bob with equal 
 probability the states $|\varphi_0\rangle$ or $|\varphi_1\rangle$.
The bit encoded by Alice is $j=0$ or $j=1$ depending on the $\ket{\varphi_j}$
state received by Bob. The density matrix $\rho_B$ held by Bob (or prepared by Alice in the PM scheme)
can  be written as:
\begin{equation}
\label{eq:B92densitymatrix1} 
\rho_B =\frac{|\varphi_0\rangle
\langle \varphi_0| +
|\varphi_1\rangle \langle \varphi_1|}{2}
 = \beta^2 |0_x\rangle \langle
0_x|+ \alpha^2 |1_x\rangle \langle 1_x|.
\end{equation}

To decode the information, Bob measures the incoming states in the
basis $\textbf{B}_k = \{ |\varphi_k\rangle,
|\overline{\varphi}_k\rangle \}$, $k=\{0,1\}$ where
$\ket{\overline{\varphi}_{k}}=\alpha\ket{0_{x}}-(-1)^{k}\beta\ket{1_{x}}$.
Upon obtaining the state $|\overline{\varphi}_k\rangle$, Bob
decodes Alice's bit as $j=k\oplus1$ (the symbol ``$\oplus$'' means
``addition modulo 2'') and labels the result as
\textit{conclusive}; on the contrary, upon obtaining the state
$|\varphi_k\rangle$, Bob is not able to decode Alice's bit
deterministically, and simply labels the result as
\textit{inconclusive}.

The same entangled state can be also used to perform the so called us-B92~\cite{LDT09}, where, 
with probability $1-p$, everything goes as in the standard
B92; on the other hand, with probability
$p \ll 1$, Alice prepares two additional, uninformative, states,
which are chosen as follows:
\begin{equation}\label{us-states}
    |us_1\rangle=|0_x\rangle~;~~~|us_2\rangle=|1_x\rangle.
\end{equation}
In particular, the states $|0_x\rangle$ and $|1_x\rangle$ are
prepared with probabilities {{$p\times\beta^2$ and
$p\times\alpha^2$}} respectively. This is necessary to assure that
Eve cannot discriminate between the density matrix pertaining to
the signal states or to the uninformative states { Eq.~\eqref{eq:B92densitymatrix1}}.
If Alice measures along the $\textbf{X}$ basis the entangled state \eqref{eq:entangled}, she will project
Bob's state in either $|0_x\rangle$, with probability $\beta^2$,
or $|1_x\rangle$, with probability $\alpha^2${ , thus}
preparing the uninformative states of the us-B92 protocol,
Eq.~\eqref{us-states}, with the correct probabilities. So, the
ent-B92 reduces to the us-B92 if one considers that the results
from Alice's $\textbf{Z}$ basis measurements are used as bits of
the final secret key, while those from the $\textbf{X}$ basis
measurements are used to perform a test against the USD attack, as
in the us-B92.
%
%
Bob's measurement remains the same as in the standard B92.
However, the presence of the uninformative states allow the users
to detect a possible USD attack~\cite{LDT09}.

{ To prove the security of the ent-B92 we} follow the approach
of device independent (DI) security proof~\cite{MPA11},
by establishing a connection between the ent-B92 protocol
and a particular Bell inequality.

Let us write the following Bell inequality, which was first introduced by
Clauser and Horne~\cite{CH74} { (``CH inequality'' for
short)}:
\begin{equation} \label{eq:CHineq}
\begin{split}
S_{CH} = P(a_1,b_1)+P(a_0,b_1)+P(a_1,b_0)\\
- P(a_0,b_0)-P(a_1)-P(b_1)\leq 0.
\end{split}
\end{equation}
Here $P(a_i,b_j)$ is the joint probability that Alice detects the
state $\vert a_i\rangle$ and Bob detects the state $\vert
b_i\rangle$, while $P(a_1)$ and $P(b_1)$ are the probabilities
that Alice and Bob respectively measure $\ket{a_1}$ and
$\ket{b_1}$, regardless of what is measured by the other user.
{This takes into account also those instances in which one of the
users receives a vacuum count. 
For instance, the term $P(a_1)$ includes the probability $P(a_1,b_v)$ that Alice measures the state $\ket{a_1}$ and Bob measures a vacuum. 
If both Alice and Bob detect a vacuum count, the event doesn't contribute to the CH inequality.}
Local realism is
verified until the above inequality is true. On the contrary,
quantum mechanics is expected to violate such an inequality in
some range of values. 

\begin{figure}[t]
  \includegraphics[width=9cm]{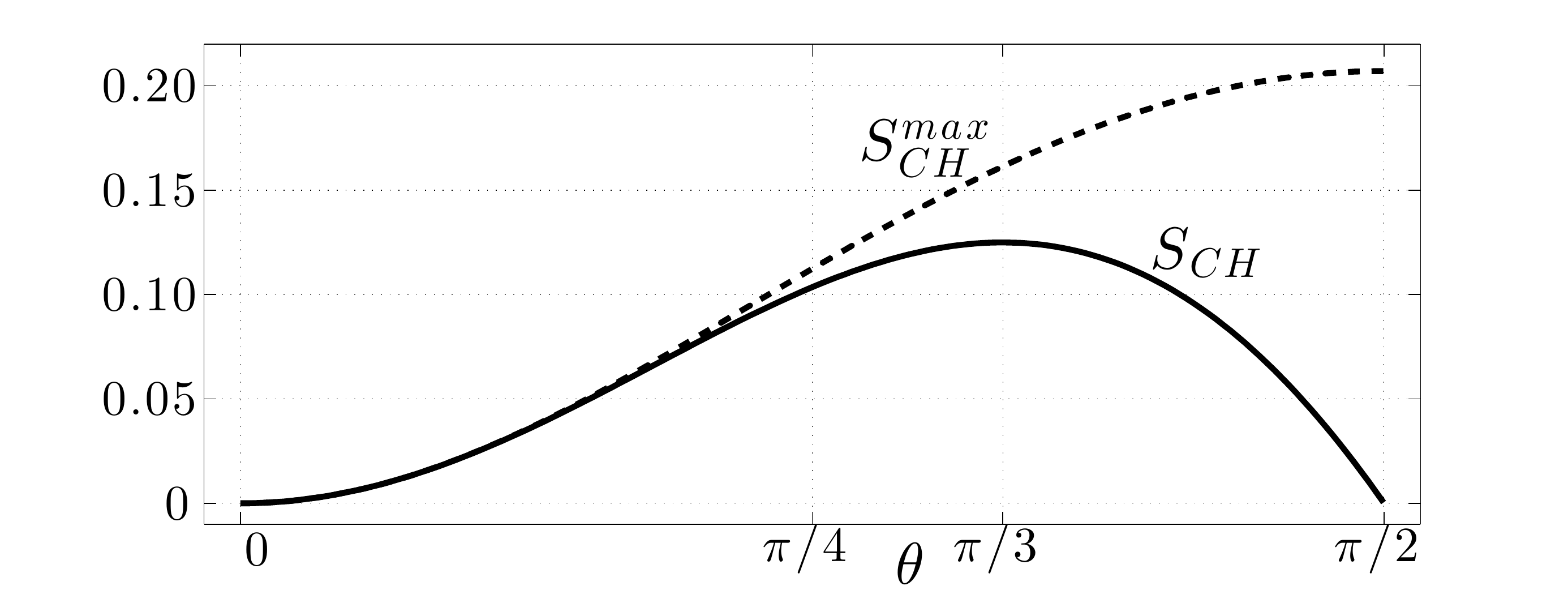}
  \caption{Plot of { $S_{CH}$} (solid line) as function of $\theta $. We also plot with dashed line
  the maximum violation achievable with the non-maximally entangled
state~\eqref{eq:entangled}, namely  $S_{CH}^{max}(\theta)=\frac{1}{2}(\sqrt{\sin^2\theta+1}-1).$
The maximum violation can be obtained if 
Alice measure along the states~\eqref{a-Hardy}, and Bob
along the new states
$|b_{0}^{\prime}\rangle =|\overline{\varphi}_{0}^{\prime}\rangle$,
$|b_{1}^{\prime}\rangle =|\overline{\varphi}_{1}^{\prime}\rangle$,
where $|\overline{\varphi}_{0}^{\prime}\rangle$ and
$|\overline{\varphi}_{1}^{\prime}\rangle$ are chosen like in
Eq.~\eqref{Al-qubit} but with an angle $\theta^{\prime}$
satisfying $\tan\theta^{\prime}=\sin \theta$.}
  \label{fig:S_CH}
\end{figure}

The goal is usually to maximize the CH
violation, but, in this paper, our first aim is to connect the violation of a
Bell inequality to the ent-B92 protocol. It is then natural to
choose the states $|a_i\rangle$ and $|b_j\rangle$ among the
ent-B92 states. We select for Alice the following states:
\begin{equation} \label{a-Hardy}
|a_{0}\rangle = |0_z\rangle \,,\quad
|a_{1}\rangle=|1_x\rangle,
\end{equation}
while we choose for Bob:
\begin{equation} \label{b-Hardy}
|b_{0}\rangle =|\overline{\varphi}_{0}\rangle \,,\quad
|b_{1}\rangle=|\overline{\varphi}_{1}\rangle.
\end{equation}
Using these states
we are able to calculate the different probabilities appearing in
Eq.~\eqref{eq:CHineq} and see for which specific values of $\theta$
the {CH} inequality is violated. After some algebra, the value
of ${ S_{CH}}$ as a function of $\theta$ { is found to be}:
\begin{equation} \label{S_CH}
{ S_{CH}(\theta)}=\frac{1}{2}\cos{\theta} (1-\cos{\theta}).
\end{equation}
This quantity is plotted in Fig.~\ref{fig:S_CH} with a solid line.
It is positive for all values of $\theta $ in the open interval
$\left( 0,\pi /2\right)$, i.e. it violates the Bell inequality for
the same interval of $\theta$ in which the ent-B92 protocol is
defined. Only the extremal points $ \theta =0$ and $\theta =\pi
/2$ are excluded. The maximum violation occurs at $\theta_{max} =
\pi/3$, corresponding to $S_{CH}=1/8$. We also plot the maximum violation
$S_{CH}^{max}(\theta)$ that can be obtained with generic measurements on $ |\Phi\rangle_{AB}$~\cite{pope92pla}.

We notice that the choice $\ket{a_1}=\ket{1_x}$, which is not
present in the traditional B92 protocol and is present in the
us-B92 protocol only for detecting the USD attack, here derives 
directly from the Bell inequality.
In fact, after choosing $\ket{a_0}$, $\ket{b_0}$ and $\ket{b_1}$
as in the B92 protocol, \eqref{a-Hardy} and \eqref{b-Hardy}, the
choice of $\ket{a_1}=\ket{1_x}$ is the one which maximizes the
violation of the CH inequality.


\section{Security proof of the \lowercase{ent}-B92}
\label{sec:4_uncsec}

{In this section we exploit a recent work by
Masanes~\textit{\underline{et al.}}~\cite{MPA11} to demonstrate
the unconditional security of the newly introduced ent-B92
protocol. {Since the security proof is based on the measured correlations, the security is
assured even if the entanglement source is under the Eve's control.
Moreover, since in Ref.~\cite{MPA11} a bound for the min entropy is found, 
the security obtained for the ent-B92 protocol is composable~\cite{kon07prl, note3}.

In~\cite{MPA11} it is explicitly given the
final \textit{secure gain}~\cite{note1} $R$ of a QKD protocol
which uses a Bell inequality in the form of a CHSH
inequality~\cite{CHSH69} to guarantee the overall security of the
transmission.
{The length $r=n_{conc}\times R$ of the secret key obtained by processing the raw key with an error-correcting protocol 
and a two-universal random function is, up to terms of order $\sqrt{n_{conc}}$, lower bounded by $H_{min} (A|E)-N_{pub}$, 
where $H_{min}(A|E)$ is the min-entropy of Alice's outcomes conditioned on EveÕs information on the joined Alice-Eve state, and $N_{pub}$ 
is the number of bits published by Alice in the error-correcting phase. 
The length of the public message necessary for correcting BobÕs errors is $N_{pub} = n_{conc}\times H (a|b)$, 
up to terms of order $\sqrt{n_{conc}}$.
From the violation of the CHSH inequality, Masanes {\it et al.} in~\cite{MPA11}  showed 
that the min entropy is bounded and obtained the following bound on the secure gain:}
\begin{equation}
\label{eq:DI-CHSH-rate}
R \geq -\log _{2}\left(\frac{1}{2}+\frac{1}{2}\sqrt{2-\frac{S_{CHSH}^{2}}{4}}\right) -H(a|b)
\end{equation}
where
\begin{equation}\label{eq:CHSHineq}
 S_{CHSH} = \langle A_1B_1\rangle
+\langle A_0B_1\rangle +\langle A_1B_0\rangle -\langle
A_0B_0\rangle.
\end{equation}
The correlation term are $\langle
A_iB_j\rangle=P(a_i,b_j)+P(\overline a_i,\overline b_j)-
P(\overline a_i,b_j)-P(a_i,\overline b_j)$ and $\bar a_i$ ($\bar
b_j$) { is} the state orthogonal to $\vert a_i\rangle$ ($\vert
b_j\rangle$).
The term containing the quantity
$S_{CHSH}$ is the one corresponding to the \textit{phase-error}
rate of the protocol~\cite{SP00}. It takes into account how much
privacy amplification~\cite{BBC+95} should be performed by the
users to remove any residual information from Eve's hands. The
term $H(a|b)$ is related to the error correction
procedure~\cite{BS94}, amounting to $h(\textrm{Q})$, with $h$ the
binary entropy~\cite{NC00} and $Q$ the QBER (quantum bit error
rate) measured on the quantum channel.
}

{The bit error rate for the ent-B92 is a measurable quantity of the protocol and doesn't represent any
 problem. It come out from the number of errors $n_{err}$, found
by the users during the error correction procedure, divided by the
number of conclusive events, $n_{con}$.
The phase error rate is given by a lower bound on Eve's
information as a function of the CHSH value. This is the main
result of Ref.~\cite{MPA11} that we want to apply here. 
However, the results of~\cite{MPA11} apply to the CHSH inequality and we have to relate it to the CH.
{By using $P(a_i,b_j)+P(\overline a_i,b_j)=P(b_j)$ and $P(a_i,b_j)+P(a_i,\overline b_j)=P(a_i)$
it is possible to show that the generic correlation term $\langle A_iB_j\rangle$ of $S_{CHSH}$ can be written as 
$4P(a_i,b_j)-2P(a_i)-2P(b_j)+1$.
{Note that this relation holds even if the vacuum counts are taken into account.
In fact, in this case it suffices to modify the correlation terms in order to consider
the losses. We use
the rule of considering the vacuum counts as a detection on
the orthogonal states $\ket{\overline a_i}$ and $\ket{\overline
b_j}$ (when the observable $A_i$ and $B_j$ are measured respectively). 
In this way the correlation term can be written as $\langle
A_i B_j\rangle = P(a_i,b_j) + [P(\overline a_i, \overline
b_j)+P(a_v,\overline b_j)+P(\overline a_i,b_v)+P(a_v,b_v)]
-[P(\overline a_i, b_j)+P(a_v, b_j)]-[P(a_i,\overline
b_j)+P(a_i,b_v)]$. Since
$P(a_i)=P(a_i,b_i)+P(a_i,\overline b_i)+P(a_i,b_v)$ even in this case we obtain
$\langle A_iB_j\rangle=4P(a_i, b_j)-2P(a_i)- 2P(b_j) + 1$.}

Starting from this relation it is straightforward to show that the two inequalities are related by 
$S_{CHSH} = 4S_{CH} + 2$}
and the secure gain becomes
\begin{equation} \label{ent-B92-gain}
R = 1-\log_{2}\left(1+\sqrt{1-4{ S_{CH}}-4{
S_{CH}}^{2}}\right)-h(\frac{n_{err}}{n_{con}}).
\end{equation}
We can notice that there is a one-to-one correspondence
between nonlocality and security: in fact, the above secure gain
is always negative when $S_{CH} < 0$, i.e., when the CH
inequality is no more violated, and is positive when $S_{CH}
> 0$, if $n_{err}=0$.
In analogy with the standard approach, in order to obtain
the secure rate of the ent-B92 protocol,  we have to
multiply the obtained gain by the number of conclusive events
collected by Alice and Bob:
\begin{equation} \label{ent-B92-rate}
r_{ent-B92}= n_{con} \times R.
\end{equation}
 Let us remark that this last step just concerns the \textit{efficiency} of the protocol, not its
\textit{security}, which only depends on the gain $R$ and on the
estimation of the CHSH value from the measured CH
inequality.}

\subsection{Resistance to losses}
\label{subsec:losses}

{\it Resistance to losses -} The DI { security proof adopted for the ent-B92} offers the
immense advantage { of making the protocol} independent of the
practical details of the implementation: Alice and Bob could even
purchase their devices directly from Eve, because the violation of
a Bell inequality would certify the secrecy of the transmission in
any case. 
{On the other side, this certification is based on a Bell
test, which is hardly feasible with current technology (see
however the proposals in~\cite{GPS10,CM11} and the high-efficiency
detectors reported in~\cite{LMN08,FFN+11}). The most difficult
step is to close the detection loophole, which
requires a very low global loss rate, from the light source to the
detectors.}
The maximum tolerable loss rate or, equivalently, the minimum
global efficiency $\eta_g$ required to close the detection
loophole, is a figure of merit of a given protocol: the lower
$\eta_g$, the more feasible the protocol.
{ It is known that to close the detection loophole} $\eta_g$
cannot be lower than $2/3$ ($67\%$), a result for the first time
found by Eberhard~\cite{Ebe93} by an inequality very similar to
our Eq.~\eqref{eq:CHineq}.
Here we want to quantify $\eta_g$ for the ent-B92,
 so to study its resistance against the losses of the
communication when the channel noise {is} zero.
Using the ent-B92 states given in
Eqs.~\eqref{a-Hardy},~\eqref{b-Hardy}, it is not difficult to see
that the resulting CH inequality can be written in terms of
$\theta$, $\eta_A$ (Alice total efficiency) and $\eta_B$ (Bob
total efficiency) as:
\begin{equation} \label{eq:CHineq-loss}
S_{CH}^{\prime}(\theta) = \left(  \eta_{A}-\frac{1}{2}\right)
\eta_{B}\sin^{2}\theta-\eta_{A} \sin^{2}\frac{\theta}{2}.
\end{equation}
Note that if $\eta_A=\eta_B=1$ this quantity coincides with that
of Eq.~\eqref{S_CH}.
It is interesting to detail a few particular cases related to this
result: (I) if $\eta_{A}=1$, then $S_{CH}^{\prime\prime}=\eta_{B}(\sin^{2}\theta)/2-\sin^{2}(\theta/2)$
    and local realism can be violated for $\eta_{B}>1/2$;
    (II) if $\eta_{B}=1$, then $S_{CH}^{\prime\prime} = \left( \eta_{A}-1/2 \right) \sin^{2} \theta - \eta_{A}
    \sin^{2}(\theta/2)$, and local realism can be violated for $\eta_{A}>2/3$;
    (III) if $\eta_{A}=\eta_{B}=\eta$, then $S_{CH}^{\prime\prime}=\left( \eta-1/2 \right) \eta\sin^{2}\theta-
    \eta\sin^{2}(\theta/2)$, and local realism can be violated for $\eta>3/4$.

The above case (III) corresponds to a symmetric
configuration, i.e. when Alice and Bob efficiencies are equal. In
this case $\eta_g = 3/4$, which is higher than the expected value
$2/3$, corresponding to Eberhard's result~\cite{Ebe93}. This is
due to the fact that the ent-B92 states are fixed and cannot be
further optimized. However, even though not optimal, this result
is quite interesting. In fact, due to the one-to-one
correspondence between secure gain and nonlocality, we can conclude that the ent-B92
provides a positive, model-independent, secure gain if the users'
efficiencies are higher than $75\%$. This value can be compared,
e.g., with that of Ref.~\cite{PAB+09}, where the minimum
efficiency required is $92.4\%$.

We can further improve on our result by exploiting the above
case (I), which refers to a non-symmetric configuration
(see also~\cite{CL07,BGS+07}), { and introducing an additional
assumption}. In concrete, we place the source of the entangled
state, Eq.~\eqref{eq:entangled}, in Alice's territory,
\textit{shielded against Eve's intrusion}. This is a standard
assumption in QKD, which holds for all PM protocols. So we are
going back to a PM configuration and consider the PM version of
the device-independent ent-B92 just introduced. This new situation
does not cover anymore the possibility that Alice's setup is
provided by Eve, but still covers the possible, involuntary,
calibration errors in Alice's devices; moreover it covers, of
course, the possibility that Bob's setup is provided directly by
Eve.  This can be also understood by considering that Alice devices are
\textit{trusted}, { even though this alternative view is}
slightly stronger than our assumption, since we let a certain
degree of untrustiness in Alice devices, represented by the
calibration errors.

Since Eve cannot modify the result of a Bell test by acting on
Alice's setup, we can safely assume that Alice efficiency is
$100\%$, thus falling into the above case (I), which
entails that Bob's efficiency can be as low as $50\%$ in order to
provide a positive secure gain independent of the implementation
details. This result coincides with what obtained in~\cite{MML08}
by a different approach { and resembles an
entanglement-steering scenario~\cite{Sch35}, where detectors are
asymmetrically treated by the users in the search of an
EPR-steering inequality violation~\cite{SJW+10}.

\subsection{Resistance to noise}
\label{subsec:noise}

In order to perform a fair comparison with the standard PM
protocols, we place the entangled light source very close to Alice
station. Then we consider a depolarizing channel acting on the
state going from Alice to Bob, as follows:
\begin{equation}\label{dep-chan}
    \rho \rightarrow \rho^{\prime}=(1-p)\rho+\frac{p}{3}(\sigma_x \rho \sigma_x
    +\sigma_y \rho \sigma_y+\sigma_z \rho \sigma_z),
\end{equation}
with $\sigma_i$ the usual Pauli matrices. It is straightforward to
show that the quantity $S_{CH}$ is modified by the depolarizing
channel as follows:
$ S_{CH}^{\prime}=1-\frac{4p}{3} S_{CH}-\frac{2p}{3}$,
where $S_{CH}$ is given by Eq.~\eqref{eq:CHineq}. By substituting
$S_{CH}^{\prime}$ in Eq.~\eqref{ent-B92-rate} we obtain the secure
rate of the ent-B92 as a function of the depolarizing parameter
$p$. The result, normalized by the total number of events
detected by the users, is plotted in Fig.~\ref{fig:fin-rates}.
\begin{figure}
  \includegraphics[width=7cm]{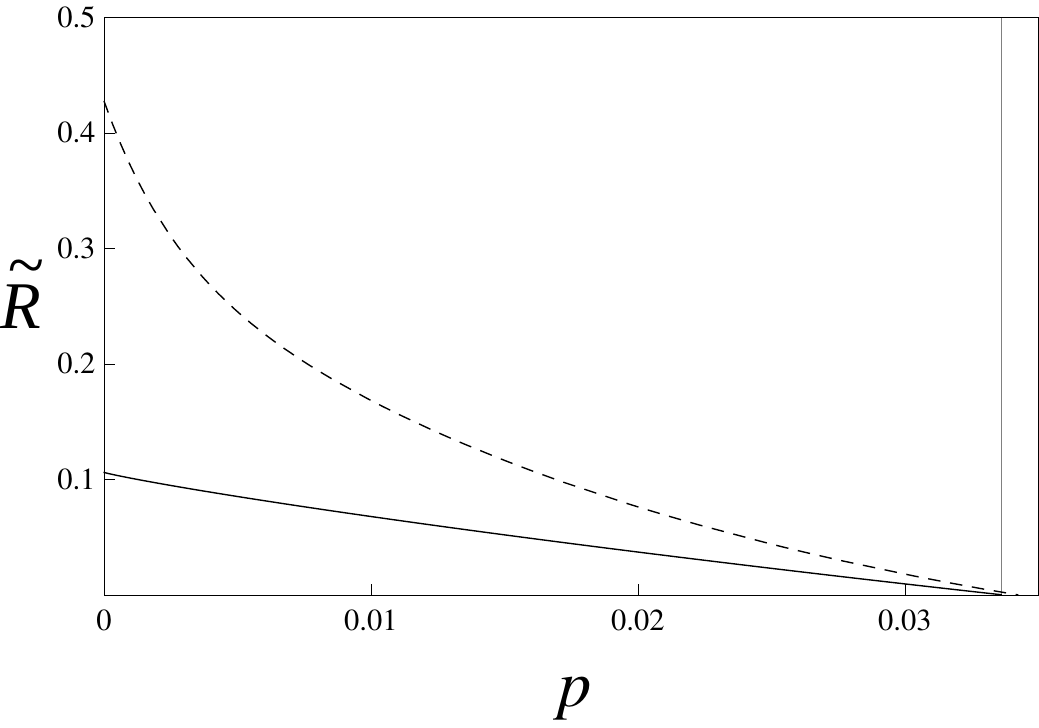}\\
  \caption{Secure { normalized rate} pertaining to the
  DI entanglement-based ent-B92 (solid line) and to PM B92 and us-B92
  (dashed line) as function of the depolarizing rate $p$ (see text).
  The maximum depolarizing rate tolerated by the ent-B92 is $p_{max}^{ent-B92}\simeq0.0336$,
  nearly the same as PM-B92, $p_{max}^{PM-B92}=0.034$~\cite{TKI03}.}
  \label{fig:fin-rates}
\end{figure}
In the same figure we also plot the secure { normalized rate
$\tilde{R}$} pertaining to the PM B92 protocols, i.e. the standard
B92 and the us-B92. In fact, they both share the same resistance
against a depolarizing channel~\cite{note2}. The ent-B92 rate
remains positive up to $p_{max}^{ent-B92}\simeq0.0336$, which is about
the same value known in the literature for PM-B92
($p_{max}^{PM-B92}=0.034$~\cite{TKI03}). In obtaining these
results, the optimal angle $\theta$ for the ent-B92 is nearly
constant regardless of the value of $p$; it is $61.56^{\circ}$ for
$p=0.01$, $62.65^{\circ}$ for $p=0.02$, $63.57^{\circ}$ for
$p=0.03$. These values of the optimal $\theta$ are quite close to
those corresponding to the optimal $\theta$ for the us-B92 (about
$55^{\circ}$)~\cite{LDT09} and for the asymmetric feedback (about
$60^{\circ}$)~\cite{LKD+10}.
It can be noted that none of the secure { rates} reported in
Fig.~\ref{fig:fin-rates} starts from the value 1. This is due to
the presence of inconclusive counts in all the B92-like protocols.
Other DI protocols provide a better efficiency and a higher
resistance to noise~\cite{MPA11,Hanggi2010}.

{ It is interesting to} observe that if the users perform
generalized measurements thus obtaining an estimate for the
quantity $S_{CH}^{max}$, this does not
improve the final secure { rate} of the ent-B92. The problem
is that in order to estimate $S_{CH}^{max}$ the user Bob has to
measure along an angle $\theta'$ which is different from the angle
$\theta$ characterizing the initial entangled state,
Eq.~\eqref{eq:entangled}. This increases the error correction
term $h(\frac{n_{err}}{ n_{con}})$ in Eq.~\eqref{ent-B92-rate},
thus reducing the overall { rate}. In fact we have verified
that the maximum tolerated noise in this case amounts to
$p\simeq0.0234$, obtained with an angle $\theta\simeq75^{\circ}$.

\subsection{Resistance to USD attack}

The security proof { of the ent-B92} is DI and covers all
possible attacks performed by Eve, included 
the USD attack, which typically represents the most dangerous menace
against B92-like protocols.
In the simplest USD attack~\cite{Ben92} Eve performs the same
measurement as Bob. When she measures
$\vert\overline{\varphi}_0\rangle$ or $\vert \overline{\varphi}
_1\rangle$ she sends $\vert \varphi_1\rangle$ or $\vert
\varphi_0\rangle$ respectively. When she measures $\vert
\varphi_0\rangle$ or $\vert \varphi_1\rangle$, she doesn't send
anything to Bob, making him detect a loss. She thus performs the
following POVM: $\Pi_1=\frac12\dens{\overline\varphi_0}$,
$\Pi_2=\frac12\dens{\overline\varphi_1}$,
$\Pi_3=\frac12\dens{\varphi_0}$, $\Pi_4=\frac12\dens{\varphi_0}$.
Depending on the measured $\Pi_i$, she sends to Bob the
corresponding following states $\ket{\chi_1}\equiv\ket{\varphi_1},
\ket{\chi_2}\equiv\ket{\varphi_0},
\ket{\chi_3}=\ket{\chi_4}\equiv\ket{vac}$, where $\ket{vac}$ is
the vacuum state.
If such an attack is brought against the quantum communication, it
alters the quantum state shared by Alice and Bob, which is no more
given by Eq.~\eqref{eq:entangled} but becomes:
$\rho'=
\sum^4_{i=1}\text{Tr}_b\Big[{\ket{\Phi}}_{ab}\bra{\Phi}\Pi_i\Big]
\otimes{\ket{\chi_i}}_b\bra{\chi_i}
$. Since the state $\rho'$ is separable it cannot violate any Bell
inequality, included the CH inequality, Eq.~\eqref{eq:CHineq},
 thus letting the users detect the attack.}
{It is worth noticing that this simple argument applies to any generic intercept-and-resend attack.}

\section{Conclusion}
\label{sec:5_conc}

In the present paper we have connected the long-standing B92
protocol to a particular form of Bell inequality. This allowed us
to provide a simple security proof for a new entanglement-based
B92-like protocol which is independent of how the QKD apparatus is
modeled.
In the proposed protocol, called "ent-B92", the same quantum
states can be used either to distill the final key or to test the
violation of a Bell inequality. Together with the fact that
only two measurement bases are required in Alice and Bob sites,
this represents a practical advantage respect to other
device-independent protocols, which have to switch between
different measurement bases in order to distill secret bits or to
perform a Bell test~\cite{ABG+07,PAB+09}. The gain and tolerance
to noise of the ent-B92 are lower than in other device-independent
protocols~\cite{MPA11,Hanggi2010}, but both the figures can be
improved by extending to the new protocol the same techniques
available for the prepare-and-measure B92~\cite{LDT09}.
Furthermore the minimum required efficiency to run the ent-B92 is
$75\%$ if the users' efficiencies are assumed to be equal, and
$50\%$ if the source of entangled states is enclosed in Alice's
territory. For other protocols~\cite{PAB+09} a value of at least
$92.4\%$ was necessary.

{

The ent-B92 protocol turns out to be a particular case of an
entanglement-distribution problem. It originates from a
non-maximally entangled state which is distributed to two distant
users. As a matter of fact, this kind of entanglement gives rise
to a local non-symmetric density matrix which can be exploited by
the users to efficiently stabilize the transmission channel
without any external communication~\cite{LKD+10}. This could
represent an important resource when entanglement has to be
distributed from a satellite or on very long distances.

A test confirming the feasibility of the ent-B92 protocol can be
experimentally performed with current technology. The non-maximally
entangled state of Eq.~\eqref{eq:entangled} can be produced in
laboratory~\cite{whit99prl,cine04pra,vall07pra,vall11pra} and a
proof of principle experiment is on the way.}

%
\begin{acknowledgments}
We thank K. Tamaki for fundamental discussions  and Ll.
Masanes, S. Pironio and A. Ac\'{i}n for a feedback on their work.
Our work was supported by EU-Project CHISTERAQUASAR and by PRIN
2009 of MIUR (Ministero dell'Istruzione, dell'Universit\`{a} e
della Ricerca). M.L. was supported by the $5\tcperthousand$~grant
C.F.~81001910439. G.V. was partially supported by the
Strategic-Research-Project QUINTET of the Department of
Information Engineering, University of Padova and the
Strategic-Research-Project QUANTUMFUTURE of the University of
Padova.
\end{acknowledgments}


\begin{thebibliography}{99}
\bibitem{note2}  The only effect of the uninformative states
is to make the single-photon B92 independent of losses. As far as
the resistance of the B92 against noise over a lossless channel is
concerned, the performances of the two protocols B92 and us-B92
are exactly the same.
\bibitem{note3} The min-entropy 
can be related to the trace distance by the quantum leftover hash lemma (see e.g.\cite{toma12nco}).
\end{thebibliography}
\end{document}